\documentclass{article}  
\usepackage{breckenr2005}
\input psfig.sty
\frompage{000} \topage{000}                                              
\def\rightmark{O-TPC for Study of Helium Burning}
\def\leftmark{Moshe Gai}
\def\n16{$^{16}N$}
\def\xo16{$^{16}O$}
\def\xco2{$CO_2$}
\def\c12ag{$^{12}C(\alpha,\gamma)^{16}O$}
\def\go16{$^{16}O(\gamma,\alpha)^{12}C$}
\def\sE1{$S_{E1}(300)$}
\def\xsE2{$S_{E2}(300)$} 

\title{Optical Readout Time Projection Chamber (O-TPC) 
for a Study of Oxygen Formation In Stellar Helium Burning  
\footnote{Work Supported by USDOE Grant No. DE-FG02-94ER40870 and the 
Yale-Weizmann Collaboration, American Committe on Weizmann 
Institute of Science.}}

\authors{
{Moshe Gai$^1$, Amos Breskin$^2$, Rachel Chechik$^2$, Volker Dangendorf$^3$, 
and Henry R. Weller$^4$.%
}\\[2.812mm]
{\normalsize
\hspace*{-8pt}$^1$ Dept. of Physics, Yale University, New Haven, CT 06520, USA.\\[0.2ex] 
\hspace*{-8pt}$^2$ Dept. of Particle Physics, Weizmann Institute of Science, Rehovot, Israel.\\
\hspace*{-8pt}$^3$ Physikalisch Technische Bundesanstalt, Braunschweig, Germany.\\
\hspace*{-8pt}$^4$ Dept. of Physics, Duke University, Durham, NC 27708, USA.
}}

\abstract{
We are developing an Optical Readout Time Projection Chamber (O-TPC) 
 detector for the study of the \c12ag reaction that determines the ratio of 
 carbon to oxygen in helium burning. This ratio is crucial for 
 understanding the final fate of a progenitor star and the nucleosynthesis 
 of elements prior to a Type II supernova; an oxygen rich star is predicted to collapse to 
 a black hole, and a carbon rich star to a neutron star. Type Ia supernovae (SNeIa) 
 are used as standard candles for measuring cosmological distances with the use 
 of an empirical light curve-luminosity stretching factor.  It is essential to understand 
 helium burning that yields the carbon/oxygen white dwarf and thus the initial stage 
 of SNeIa.  The O-TPC is intended for use with high intensity photon beams extracted 
 from the HI$\gamma$S/TUNL facility at Duke University to study the \go16 reaction, 
 and thus the direct reaction at energies as low as 0.7 MeV. We are 
 conducting a systematical study of the best oxygen containing gas with light emitting 
 admixture(s) for use in such an O-TPC. Preliminary results with $CO_2$ + TEA mixture 
 were obtained.}



\begin{document}
 
\maketitle
\setcounter{page}{1}

\section{Introduction: Oxygen Formation in Helium 
  Burning and The \c12ag Reaction}

Carbon and oxygen are produced during helium burning in Red Giant stars 
before they undergo supernova explosions.  These are some of the 
most important elements required to support life as we know it on earth, and as such, 
the understanding of the origin of carbon and oxygen was designated by Willie Fowler 
in his 1984 Nobel talk, the Òholy grailÓ of Stellar Nuclear Physics \cite{Fo84}.  Moreover, 
the ratio of carbon to oxygen (C/O) at the end of helium burning is one of the most 
important parameters for understanding stellar evolution.  In a massive star (at least 
8M$_\odot$) it determines whether the star that undergoes Type II supernova, collapses 
to a black hole or a neutron star \cite{We93}.  A sun-like star that ends up as a carbon 
plus oxygen white dwarf, is the progenitor star for a Type Ia supernova explosion (SNeIa), 
that are now used as a standard candle for measuring distances comparable to the size 
of the observed universe (13.7 Billion Light Years) \cite{Phil}.  These Hubble type 
measurements of cosmological distances (using SNeIa) allow us to conclude that the 
universe is composed for the most part of dark matter (approximately 23\%), and dark 
energy (approximately 73\%) with the latter giving rise to a recent (5 Billion years ago) 
accelerated expansion of the universe \cite{Perl}. Thus the understanding of stellar 
evolution and the calibration of peak luminosity of SNeIa is essential for placing this 
conclusion on firm theoretical foundation. The C/O ratio of a white dwarf affects the 
peak luminosity and the shape of the light curve of SNeIa \cite{SNIa}. This ratio must 
be determined from laboratory measurements.

The outcome of helium burning is the formation of the two elements, carbon and oxygen 
\cite{Fo84,We93,Ga99}. The first stage in helium burning, the formation of $^{12}C$ via the 
triple alpha-particle capture reaction, is well understood \cite{Fo84} and is denoted, 
using standard Nuclear Physics notation, as the $^8Be(\alpha,\gamma)^{12}C$ 
reaction. Thus one must extract the cross section for forming oxygen via the fusion of 
carbon plus helium denoted by \c12ag. This reaction is dominated by two partial waves 
and one must extract the p-wave (S$_{E1}$) and d-wave (S$_{E2}$) astrophysical cross section 
factors for the reaction as discussed in \cite{Ga99}.  The cross section factors must be known at 
the Gamow peak (300 keV) relevant for stellar environment, with high accuracy of approximately 
10\% or better. Current data are measured at energies not lower than 1.2 MeV and the 
extrapolation to stellar burning energies of 0.3 MeV is particularly difficult due to substantial 
contributions from bound states of $^{16}O$ at very low energies.

\subsection{Beta-Delayed Alpha-Particle Emission of \n16}

The cross section of the fusion reaction is written in terms of an astrophysical cross section factor 
(the S-Factor) times a kinematical factor that is due to the small penetrability of the two interacting 
charged particles. During the 1990Õs it was suggested that one can measure the p-wave E1 S-factor 
using the beta-decay of $^{16}N$ \cite{Zh93,Zh93a,Bu93,Az94,Fr96,Fr96a} followed by alpha decay 
of $^{16}O$.  This theoretical concept of using the beta-decay of $^{16}N$ was introduced to circumvent 
the fast drop of the cross section at lower energies due to penetration through the Coulomb barrier.  
Indeed some used the $^{16}N$ data to quote the S-factor with 25\% uncertainty \cite{Az94} and the 
problem was considered solved.  But later it was shown by Hale \cite{Ha96} that the interpretation of the 
experimental data on the beta-decay of $^{16}N$ is inconclusive and the astrophysical cross section 
factor is ill determined by at least a factor of 4 and as much as 8. A recent measurement at lower energies \cite{Hammer01} suggests a d-wave cross section factor that is at least twice larger than 
"the accepted value", and the their low energy data point(s) measured with low precision 
can not rule out a small p-wave cross section factor.

We conclude that the astrophysical cross section factor must be measured directly at energies as low as 
possible and as close as possible to the Gamow window (300 keV) as it is in stellar helium burning.  
A unique opportunity \cite{HIGS,Phys} presented itself with the newly constructed High Intensity Gamma 
Source (HI$\gamma$S) now operating at Duke University, as we discuss below.

\section{The Proposed \go16 Experiments}

\centerline{\psfig{figure=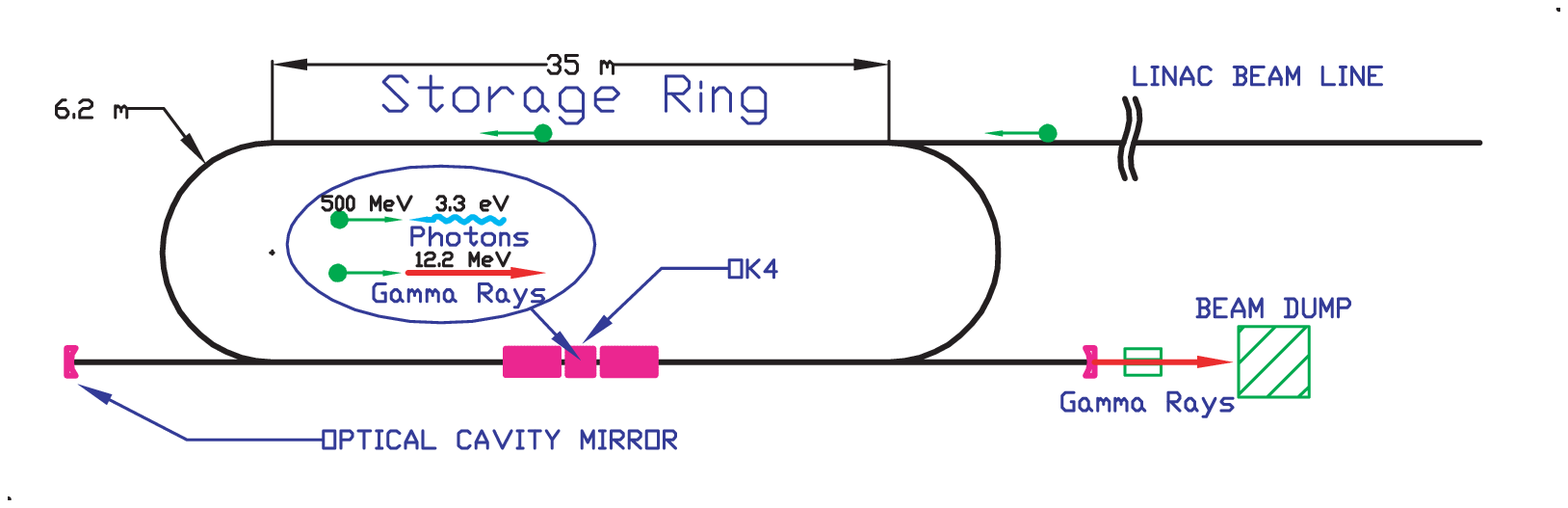,height=1.7in}}

\begin{center}
\underline{Fig. 1:}  A schematic diagram of the newly constructed HI$\gamma$S 
facility of TUNL at Duke University \cite{HIGS}.
\end{center}

In order to determine the cross section of the \c12ag reaction at relative energies as low as 700 KeV,
considerably lower than measured till now, it is useful to have an experimental setup with three 
conditions: enhanced  cross section, high luminosity and low background.  It turns out that the 
use of the inverse process, the \go16 reaction may indeed satisfy all three conditions. The cross 
section of the \go16 inverse reaction (with polarized photons) at the kinematical region of interest 
(photons approx 8-10 MeV) is larger by a factor of 80-160 than the cross section of the direct \c12ag 
reaction. Note that the 100\% linearly polarized gamma-ray beam (available at
HI$\gamma$S) yields an extra extra factor of two in the enhancement. It is evident 
that with similar luminosities and lower background (see below), the photodissociation cross section 
can be measured at center of mass energies as low as 700 keV, where the direct \c12ag  cross section 
can be estimated to be of the order of 10 pb (thus the corresponding 
photodissociation cross section is approximately 1nb). A very 
small contribution (less than 5\%) from cascade gamma decay can not be measured in this experiment, 
but appears to be negligible  and below the design goal accuracy of our measurement of $\pm 10\%$.

The High Intensity Gamma Source (HI$\gamma$S) \cite{HIGS}, shown in Fig. 1, has already achieved 
many milestones and is  approaching its design goals for 2-200 MeV gammas. This experiment will 
place a stringent demand of 8-10 MeV gammas at a resolution of 0.5\% or better and intensity of order 
$10^9$ /sec. The backscattered photons of the HI$\gamma$S facility will be collimated and will enter the 
target/detector Optical Time Projection Chamber (O-TPC) setup as we propose below. With a Q value 
of -7.162 MeV, our experiment will utilize gammas of energies ranging from 7.9 to 10 MeV.  Note that 
the emitted photons will be linearly polarized and the 
emitted particles will be primarily in a horizontal plane 
with a $sin^2\phi$ azimuthal angular dependence. This simplifies the tracking of particles in this 
experiment. The pulsed photon beam (0.1 nsec every 180 nsec with at most 500 gammas per pulse) 
provides a trigger for the track-recording image-intensified cooled CCD camera of the O-TPC; the 
scattering angle will be deduced from the reconstructed tracks relative to the beam direction with high
accuracy using the (8 cm long) alpha tracks and (2 cm long) carbon tracks. The time projection information from the O-TPC will yield the azimuthal angle of the event of interest. 
Background events will be discriminated with time-of-flight techniques, and gating of the CCD 
camera. The image intensified CCD camera will be triggered by light detected in the PMT, see below. 
Background events (mainly) from oxygen isotopes, will be discriminated 
using the TPC as a calorimeter with a 5\% energy resolution. Time of flight techniques, and flushing
of the CCD between two events will also be used. To reduce noise, the CCD will be cooled. We note 
that similar research program with high intensity photon beams and a TPC already exists at the 
RCNP at Osaka, Japan \cite{Shima}, proving that tracks from low energy light ions can be identified 
in the TPC with a manageable electron background resulting from the intense gamma beams. We 
also performed a test using silicon detectors in helium gas exposed to 25 MeV photons and measured 
the resulting electron background. We conclude that this background is manageable.

\subsection{Proposed Time Projection Chamber (TPC)}

We are developing an Optical Readout Time Projection Chamber (TPC), based  
on the TPC constructed in the Physikalisch Technische Bundesanstalt,
(PTB) in Braunschweig, Germany and the Weizmann Institute, Rehovot, 
Israel \cite{NIMA}, for the detection of alphas and carbon, the byproduct 
of the photodissociation of \xo16.  Since the range of available alphas is 
approximately 8 cm (at 100 mbars) the TPC is designed to be 40 cm wide and up to 
one meter long. We plan to first construct a 40 cm long TPC for 
initial use at the HI$\gamma$S beam line at TUNL/Duke. The TPC 
is largely insensitive to single Compton electrons and it  
allows for tracking of both alphas and carbons 
emitted back to back from the beam position 
in time correlation. The very different range of alphas and carbons (approximately 
a factor of 4), and differences in the lateral ionization
density, will aid us in particle identification. The TPC will also allow us to 
measure angular distributions with respect to the photon beam  
thus separating the E1 and E2 components of the \c12ag reaction. The 
energy resolution of the TPC (5\% or better) will allow us to exclude 
events from the photodissociation of nuclei other than $^{16}O$, including 
isotopes of oxygen, that are present in the gas.  
In Fig. 2, based on Titt {\em et al.} \cite{NIMA}, we show 
a schematic diagram of the proposed Optical Readout TPC. \\

\centerline{\psfig{figure=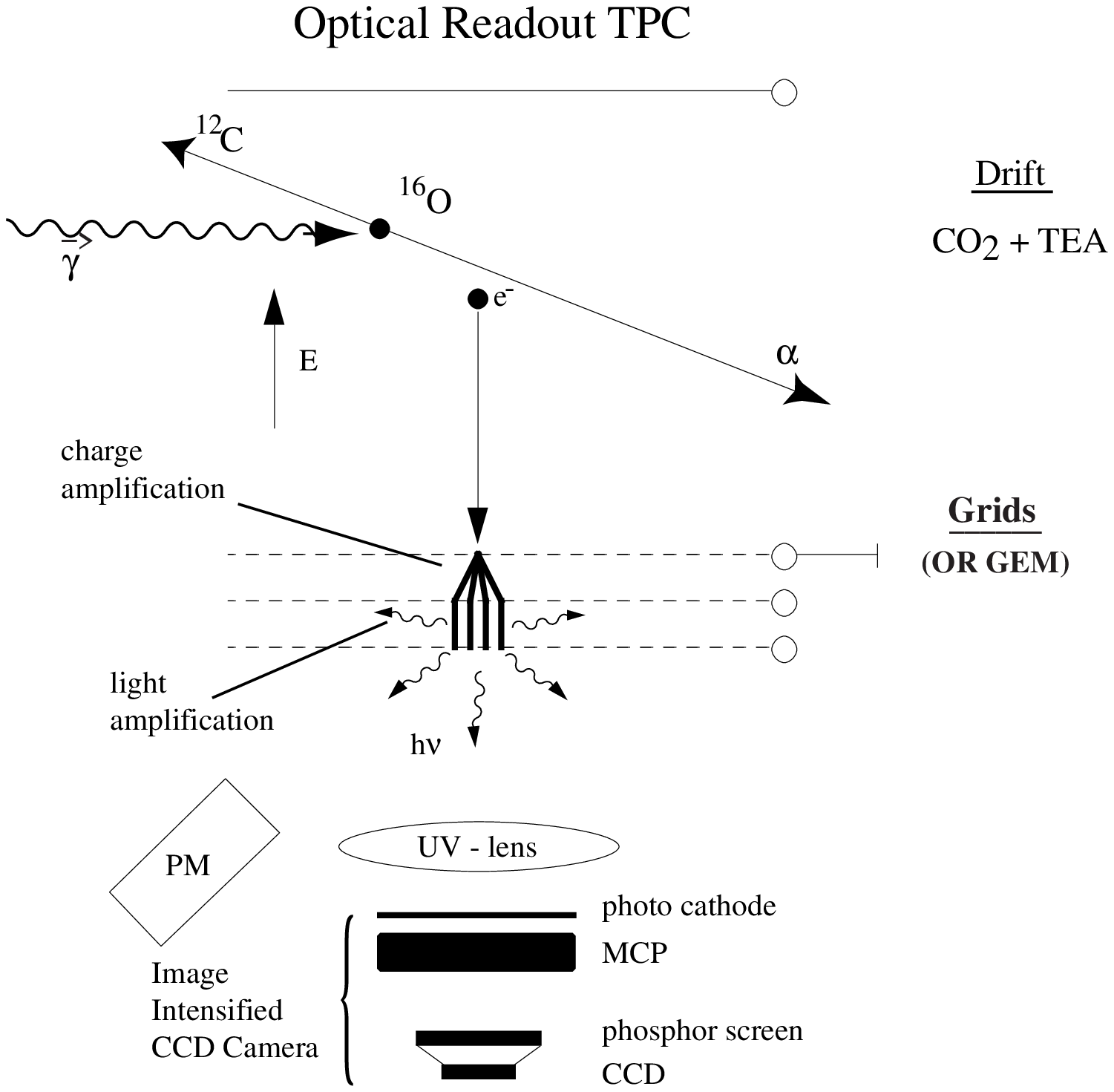,height=3.5in}}

\begin{center}
\underline{Fig. 2:} A schematic diagram of the O-TPC that we propose to construct 
for this study. It is designed following the O-TPC used by the Weizmann-PTB 
collaboration \cite{NIMA}.
\end{center}

The  photon beam will enter the TPC through an 
entrance window in the drift chamber part of 
the TPC and mainly produce background $e^+e^-$ pairs and a smaller 
amount of Compton electrons, as well as the photodissociation 
of various nuclei present in the $CO_2$ + TEA gas mixture, including 
$^{16}O$. The charged particles resulting from the photodissociation process will induce 
secondary ionization electrons in the gas that drift along the electric field. The drift times 
are in the $\mu$sec range. The electrons that will reach electron-amplification 
element, e.g. cascaded parallel-grid avalanche multipliers \cite{NIM} or the recently 
developed Gas Electron Multipliers (GEM) \cite{NIM1,NIM2}, will be multiplied in an avalanche 
process by approximately a factor of 10$^5$, yielding charge signals. 
Scintillation processes induced by avalanche-electron excitations of 
selected gas molecules, e.g. triethylamine (TEA) \cite{NIMA}  
yield copious amount of UV photons. Optically imaged cascaded GEMs and the Thick GEM-like 
(T-GEM) multipliers recently developed at the Weizmann Institute \cite{NIM3}, have higher 
multiplication factors and are expected to provide higher photon yields. A fraction of the avalanche 
light will be detected by the photomultiplier (PM) tube (Fig. 2). The PMT signal together with 
various grid signal(s), see Fig. 2, will be used 
in the trigger configuration of the Image Intensifier and the 
CCD camera, which takes a picture of the tracks. Dedicated pattern recognition algorithms will  
select the typical back-to-back Alpha-Carbon tracks. The background electrons 
lose approximately 100 keV 
in the entire TPC and and will be removed by appropriate thresholds. Events from the 
photodissociation of nuclei other than $^{16}O$ will be removed by measuring 
the total energy (Q-value) of the event with a resolution of approximately 5\%.

We have conducted light output tests with $CO_2$ + TEA gas mixture and find promising 
results for electron amplification and light output in this gas mixture. This work is in progress 
at the Weizmann Institute.

\section{Design Goal}

\centerline{\psfig{figure=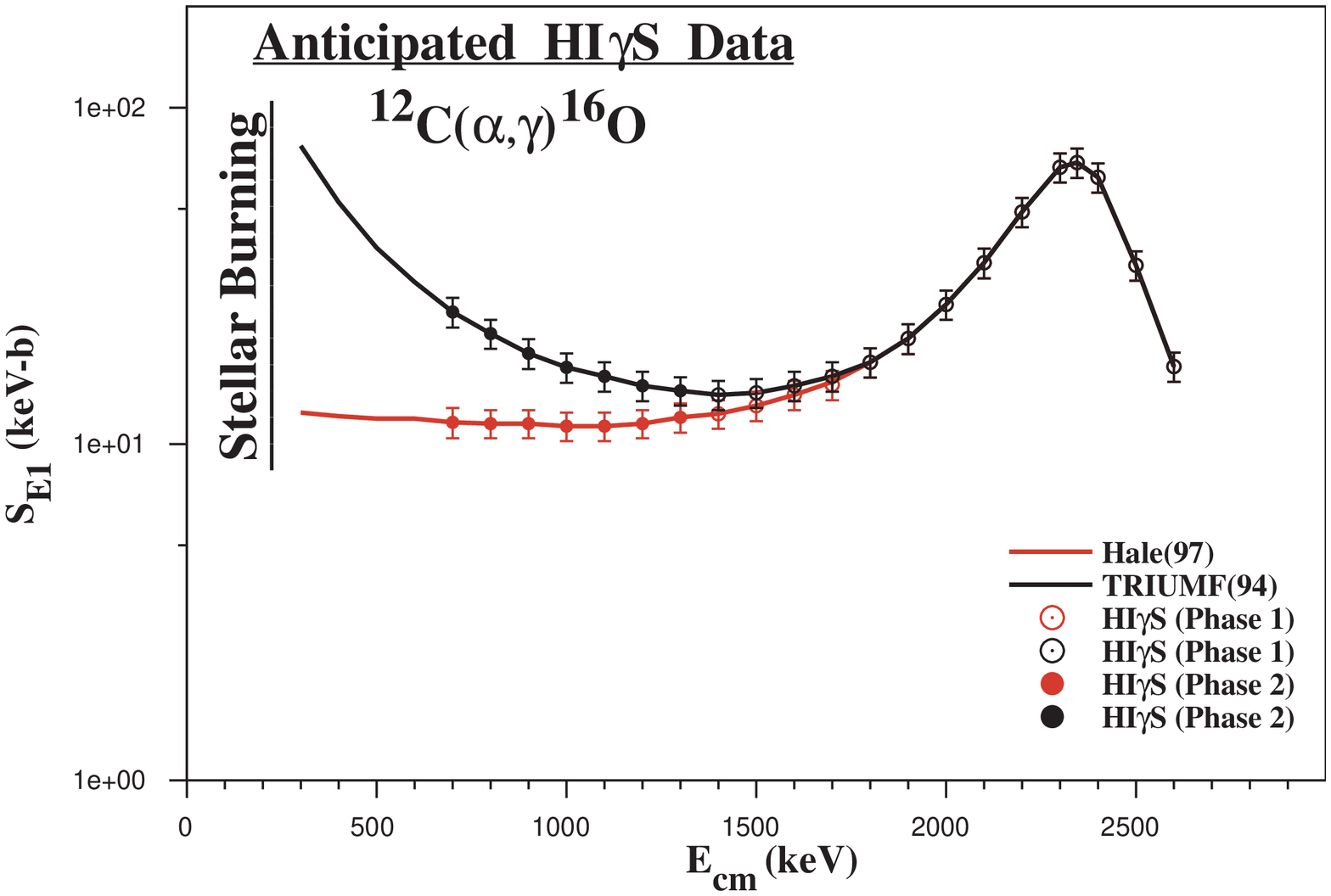,height=3.3in}}

\begin{center}
\underline{Fig. 3:} Anticipated results for the p-wave astrophysical cross section 
factor compared with predictions.
\end{center}

The time projection of the drift electrons will allow us to measure the inclination angle $\phi$ 
of the plane of the byproducts, and the tracks themselves will 
allow for measurement of the scattering angle 
$\theta$, both with an accuracy of better than two degrees. The so obtained angular distributions will  
be fitted with Legendre polynomials to provide the contributions of the 
p- and d- partial waves, from which 
the cross section factors will be deduced. The results of our simulation showing the anticipated
astrophysical S-factors for p-waves (SE1) and d-waves (SE2) are shown in 
Figs. 3 and 4, respectively\\
\     \\
\centerline{\psfig{figure=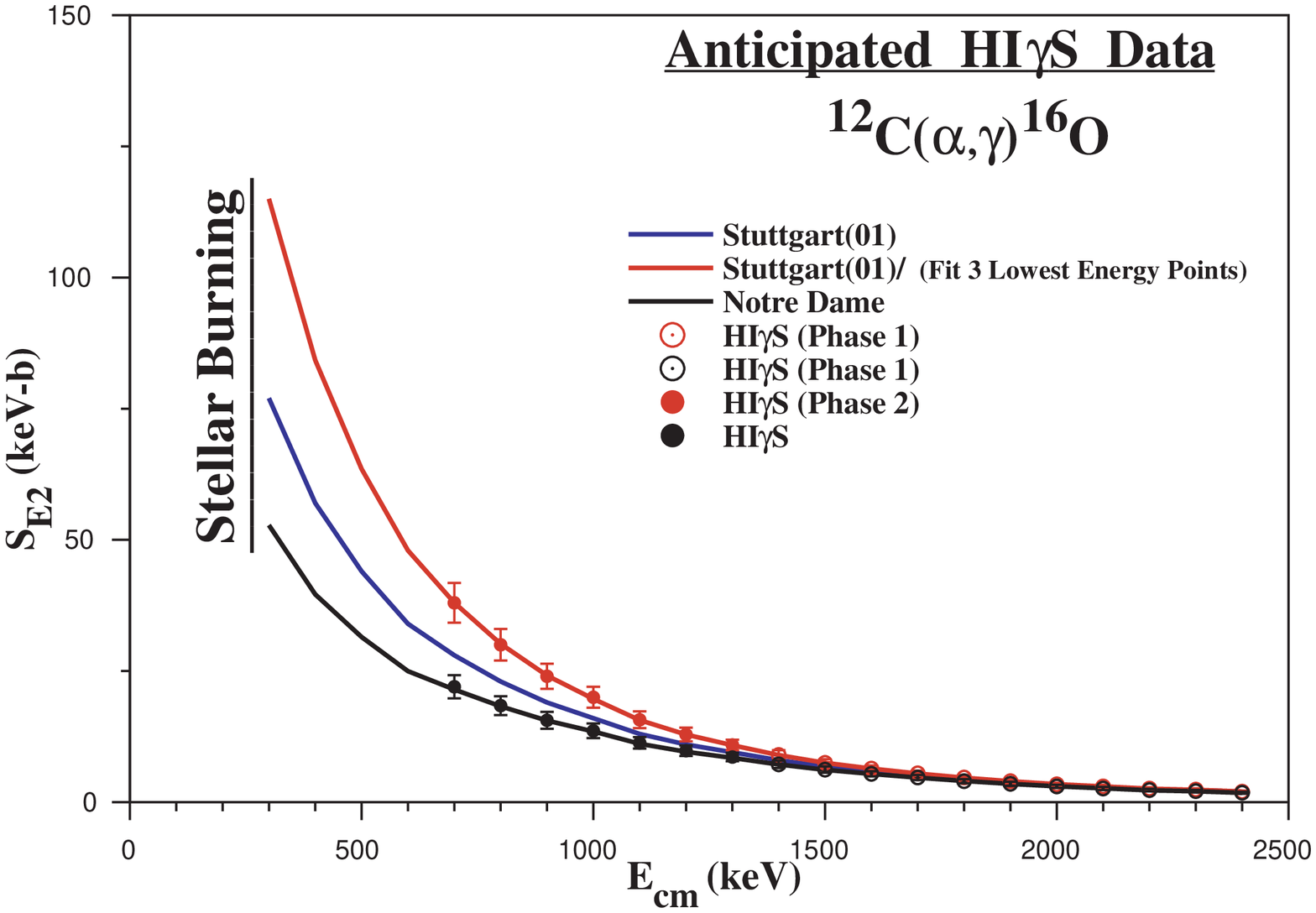,height=3.5in}}

\begin{center} 
\underline{Fig. 4:} Anticipated results for the d-wave astrophysical cross section 
factor compared with predictions.
\end{center}

The luminosity of our proposed \go16 experiment 
can be very large. For example, with a 30 cm long fiducial length target with 
$CO_2$ at a pressure of 76 torr (100 mbar) and a photon beam of $1 \times 
10^9$ /sec, we obtain a luminosity of $1.5 \times 10^{29}\ sec^{-1}cm^{-2}$
(15 nb$^{-1}$/day).  Thus a measurement of the 
photodissociation of $^{16}O$ with a cross section of 1 nb, 
yields 15 counts per day, leading to a design goal sensitivity for measuring the direct 
\c12ag reaction with a cross section as low as 10 pb, corresponding to energies 
as low as 700 keV. A mark I experiment to measure coincidences between 
$\alpha$-particles and $^{12}C$ is in progress at the TUNL/HI$\gamma$S facility.

\end{document}